\documentclass[aps,print,showpacs,a4paper,12pt,onecolumn]{revtex4}
\usepackage{amssymb}
\usepackage{amsmath}
\usepackage{graphicx}

\input{tcilatex}

\begin{document}

\title{Magnetic Soliton and Soliton Collisions of Spinor Bose-Einstein
Condensates in an Optical Lattice}
\author{Z. D. Li$^{1}$, P. B. He$^{2}$, L. Li$^{1}$, J. Q. Liang$^{1}$, and
W. M. Liu$^{2}$}
\affiliation{$^{1}$Institute of Theoretical Physics and Department of Physics, Shanxi
University, Taiyuan 030006, China\\
$^{2}$Joint Laboratory of Advanced Technology in Measurements, Beijing
National Laboratory for Condensed Matter Physics, Institute of Physics,
Chinese Academy of Sciences, Beijing 100080, China}

\begin{abstract}
We study the magnetic soliton dynamics of spinor Bose-Einstein condensates
in an optical lattice which results in an effective Hamiltonian of
anisotropic pseudospin chain. A modified Landau-Lifshitz equation is derived
and exact magnetic soliton solutions are obtained analytically. Our results
show that the time-oscillation of the soliton size can be controlled in
practical experiment by adjusting of\textbf{\ }the light-induced
dipole-dipole interaction. Moreover, the elastic collision of two solitons
is investigated.
\end{abstract}

\pacs{03.75.Lm, 05.30.Jp, 67.40.Fd}
\maketitle

\section{Introduction}

Recently, spinor Bose-Einstein condensates (BECs) trapped in optical
potentials have received much attention in both experimental \cite%
{Stenger,Anderson,Liu} and theoretical studies \cite{Ho}. Spinor BECs have
internal degrees of freedom due to the hyperfine spin of the atoms which
liberate a rich variety of phenomena such as spin domains \cite{Miesner} and
textures \cite{Ohmi}. When the potential valley is so deep that the
individual sites are mutually independent, spinor BECs at each lattice site
behave like spin magnets and can interact with each other through both the
light-induced and the static, magnetic dipole-dipole interactions. These
site-to-site dipolar interactions can cause the ferromagnetic phase
transition \cite{Pu,Kevin} leading to a \textquotedblleft
macroscopic\textquotedblright\ magnetization of the condensate array and the
spin-wave like excitation \cite{Pu,Zhang} analogous to the spin-wave in a
ferromagnetic spin chain. For the real spin chain, the site-to-site
interaction is caused mainly by the exchange interaction, while the
dipole-dipole interaction is negligibly small. For the spinor BECs in the
optical lattice, the exchange interaction is absent. The individual spin
magnets are coupled by the magnetic and the light-induced dipole-dipole
interactions \cite{Zhang} which are no longer negligible due to the large
number of atoms $N$ at each lattice site, typically of the order of 1000 or
more. Therefore, the spinor BECs in an optical lattice offer a totally new
environment to study spin dynamics in periodic structures. The magnetic
soliton excited by the interaction between the spin waves \cite{Zhang} is an
important and interesting phenomenon in spinor BECs. In this paper, we
demonstrate that the magnetic soliton and elastic soliton collision are
admitted for spinor BECs in a one-dimensional optical lattice and are
controllable by adjusting of the light-induced and the magnetic
dipole-dipole interactions.

The Heisenberg model of spin-spin interactions is considered as the starting
point for understanding many complex magnetic structures in solids. In
particular, it explains the existence of ferromagnetism and
antiferromagnetism at temperatures below the Curie temperature. The magnetic
soliton \cite{Kosevich}, which describes localized magnetization, is an
important excitation in the Heisenberg spin chain \cite%
{Tjon,Li,Ablowitz,Huang}. The Haldane gap \cite{Haldane} of antiferromagnets
has been reported in integer Heisenberg spin chain. By means of the neutron
inelastic scattering \cite{Kjems78} and electron spin resonance \cite%
{Asano00}, the magnetic soliton has already been probed experimentally in
quasi-one dimensional magnetic systems. Solitons can travel over long
distances with neither attenuation nor change of shape, since the dispersion
is compensated by nonlinear effects. The study of soliton has been conducted
in as diverse fields as particle physics, molecular biology, geology,
oceanography, astrophysics, and nonlinear optics. Perhaps the most prominent
application of solitons is in high-rate telecommunications with optical
fibers. However, the generation of controllable solitons is an extremely
difficult task due to the complexity of the conventional magnetic materials.
The spinor BECs seems an ideal system to serve as a new test ground for
studying the nonlinear excitations of spin waves both theoretically and
experimentally.

The outline of this paper is organized as follows: In Sec. II the
Landau-Lifshitz equation of spinor BEC in an optical lattice is derived in
detail. Next, we obtain the one-soliton solution of spinor BEC in an optical
lattice. The result shows that the time-oscillation of the amplitude and the
size of soliton can be controlled by adjusting of the light-induced
dipole-dipole interaction. We also present that the magnetization varies
with time periodically. In Sec. VI, the general two-soliton solution for
spinor BEC in an optical lattice is investigated. Analysis reveals that
elastic soliton collision occurs and there is a phase exchange during
collision. Finally, our concluding remarks are given in Sec. V.

\section{Landau-Lifshitz equation of spinor BEC in an optical lattice}

The dynamics of spinor BECs trapped in an optical lattice\textbf{\ }is
primarily governed by three types of two-body interactions: spin-dependent
collision characterized by the $s$-wave scattering length, magnetic
dipole-dipole interaction (of the order of Bohr magneton $\mu _{B}$), and
light-induced dipole-dipole interaction adjusted by the laser frequency in
experiment. Our starting point is the Hamiltonian describing an $F=1$ spinor
condensate at zero temperature trapped in an optical lattice, which is
subject to the magnetic and the light-induced dipole-dipole interactions and
is coupled to an external magnetic field via the magnetic dipole Hamiltonian 
$H_{B}$ \cite{Ho,Miesner,Ohmi,Pu}, 
\begin{eqnarray}
H &=&\sum_{\alpha }\int d\mathbf{r}\hat{\psi}_{\alpha }^{\dagger }(\mathbf{r}%
)[-\frac{\hbar ^{2}\nabla ^{2}}{2m}+U_{L}(\mathbf{r})]\hat{\psi}_{\alpha }(%
\mathbf{r})  \notag \\
&&+\sum_{\alpha ,\beta ,\upsilon ,\tau }\int d\mathbf{r}d\mathbf{r}^{\prime }%
\hat{\psi}_{\alpha }^{\dagger }(\mathbf{r})\hat{\psi}_{\beta }^{\dagger }(%
\mathbf{r}^{\prime })\left[ {}\right. U_{\alpha \upsilon \beta \tau }^{coll}(%
\mathbf{r,r}^{\prime })+U_{\alpha \upsilon \beta \tau }^{d-d}(\mathbf{r,r}%
^{\prime })\left. {}\right] \hat{\psi}_{\tau }(\mathbf{r}^{\prime })\hat{\psi%
}_{\upsilon }(\mathbf{r})+H_{B},  \label{hamilton}
\end{eqnarray}%
where $\hat{\psi}_{\alpha }\left( r\right) $ is the field annihilation
operator for an atom in the hyperfine state $\left\vert f=1,m_{f}=\alpha
\right\rangle $, $U_{L}(\mathbf{r})$ is the lattice potential, the indices $%
\alpha ,\beta ,\upsilon ,\tau $ which run through the values $-1,0,1$ denote
the Zeeman sublevels of the ground state. The parameter $U_{\alpha \upsilon
\beta \tau }^{coll}(\mathbf{r,r}^{\prime })$ describes the two-body
ground-state collisions and $U_{\alpha \upsilon \beta \tau }^{d-d}(\mathbf{%
r,r}^{\prime })$ includes the magnetic dipole-dipole interaction and the
light-induced dipole-dipole interaction.

When the optical lattice potential is deep enough there is no spatial
overlap between the condensates at different lattice sites. We can then
expand the atomic field operator as $\hat{\psi}\left( \mathbf{r}\right)
=\tsum\nolimits_{n}$ $\tsum\nolimits_{\alpha =0,\pm 1}\hat{a}_{\alpha
}\left( n\right) \phi _{n}\left( \mathbf{r}\right) $, where $n$ labels the
lattice sites, $\phi _{n}(\mathbf{r})$ is the condensate wave function for
the $n$th microtrap and the operators $\hat{a}_{\alpha }(n)$ satisfy the
bosonic commutation relations $[\hat{a}_{\alpha }(n),\hat{a}_{\beta }^{\dag
}(l)]=\delta _{\alpha \beta }\delta _{nl}$. It is assumed that all Zeeman
components share the same spatial wave function. If the condensates at each
lattice site contain the same\textbf{\ }number of atoms\textbf{\ }$N$, the
ground-state wave functions for different sites have the same form $\phi
_{n}\left( \mathbf{r}\right) =\phi _{n}\left( \mathbf{r}-\mathbf{r}%
_{n}\right) $.

In this paper we consider a one-dimensional optical lattice along the $z$%
-direction, which we also choose as the quantization axis. In the absence of
spatial overlap between individual condensates, and neglecting unimportant
constants, we can construct the effective spin Hamiltonian \cite{Zhang,Pu}
as 
\begin{equation}
H=\sum_{n}[\lambda _{a}^{\prime }\mathbf{\hat{S}}_{n}^{2}-\sum_{l\neq
n}\lambda _{nl}\mathbf{S}_{n}\cdot \mathbf{S}_{l}+2\sum_{l\neq n}\lambda
_{nl}^{ld}\hat{S}_{n}^{z}\hat{S}_{l}^{z}-\gamma \hat{S}_{n}\cdot \mathbf{B}],
\label{hamilton2}
\end{equation}
where $\lambda _{nl}=2\lambda _{nl}^{ld}+\lambda _{nl}^{md}$, $\lambda
_{nl}^{md}$ and $\lambda _{nl}^{ld}$ represent the magnetic and the
light-induced dipole-dipole interaction respectively. The direction of the
magnetic field $\mathbf{B}$ is along the one-dimensional optical lattice and 
$\gamma =g_{F}\mu _{B}$ is the gyromagnetic ratio. The spin operators are
defined as $\mathbf{S}_{n}=\hat{a}_{\alpha }^{\dag }(n)\mathbf{F}_{\alpha
\upsilon }\hat{a}_{\upsilon }(n)$, where $\mathbf{F}$ is the vector operator
for the hyperfine spin of an atom, with components represented by $3\times 3$
matrices in the $\left\vert f=1,m_{f}=\alpha \right\rangle $ subspace. The
first term in Eq. (\ref{hamilton2})\textbf{\ }is resulted\textbf{\ }from the
spin-dependent interatomic collisions at a given site, with $\lambda
_{a}^{\prime }=(1/2)\lambda _{a}\int d^{3}r\left\vert \phi _{n}(\mathbf{r}%
)\right\vert ^{4}$, where $\lambda _{a}$ characterizes the spin-dependent $s$%
-wave collisions. The second and the third terms describe the site to site
spin coupling induced by the static magnetic field dipolar interaction and
the light-induced dipole-dipole interaction. For $\lambda _{nl}\neq 0$, the
transfer of transverse excitation from site to site is allowed, resulting in
the distortion of the ground-state spin structure. This distortion can
propagate and hence generate spin waves along the atomic spin chain. For an
optical lattice created by blue-detuned laser beams, the atoms are trapped
in the dark-field nodes of the lattice and the light-induced dipole-dipole
interaction is very small \cite{Kevin}. However, this small light-induced
dipole-dipole interaction induces the amplitude and size of the soliton
varying with time periodically as we will show in the following section.

From Hamiltonian (\ref{hamilton2}) we can derive the Heisenberg equation of
motion at $k$th site for the spin excitations. When the optical lattice is
infinitely long and the spin excitations are in the long-wavelength limit,
i.e., the continuum limit, $S_{k}\rightarrow S\left( z,t\right) $, we obtain
the Landau-Lifshitz equation of a spinor BECs in an optical lattice as
follows: 
\begin{eqnarray}
\frac{\partial S^{x}}{\partial t} &=&\frac{2\lambda }{\hbar }[a^{2}(S^{y}%
\frac{\partial ^{2}}{\partial z^{2}}S^{z}-S^{z}\frac{\partial ^{2}}{\partial
z^{2}}S^{y})-4\frac{\lambda ^{ld}}{\lambda }S^{y}S^{z}]+\frac{\gamma BS^{y}}{%
\hbar },  \notag \\
\frac{\partial S^{y}}{\partial t} &=&\frac{2\lambda }{\hbar }[a^{2}(S^{z}%
\frac{\partial ^{2}}{\partial z^{2}}S^{x}-S^{x}\frac{\partial ^{2}}{\partial
z^{2}}S^{z})+4\frac{\lambda ^{ld}}{\lambda }S^{z}S^{x}]-\frac{\gamma BS^{x}}{%
\hbar },  \notag \\
\frac{\partial S^{z}}{\partial t} &=&\frac{2\lambda }{\hbar }[a^{2}(S^{x}%
\frac{\partial ^{2}}{\partial z^{2}}S^{y}-S^{y}\frac{\partial ^{2}}{\partial
z^{2}}S^{x})],  \label{magnet}
\end{eqnarray}%
where we assume that all nearest-neighbor interactions are the same, namely $%
\lambda _{<nl>}=\lambda $, which is a good approximation in the
one-dimensional optical lattice for the large lattice constant \cite{Konotop}%
.\textbf{\ }In a rotating frame around $z$-axis with angular frequency $%
\frac{\gamma B}{\hbar }$\ the spin vector $S$\ is related with the original
one by the transformation 
\begin{equation}
S^{x}=S^{x^{\prime }}\cos (\frac{\gamma B}{\hbar }t)+S^{y^{\prime }}\sin (%
\frac{\gamma B}{\hbar }t),S^{y}=S^{y^{\prime }}\cos (\frac{\gamma B}{\hbar }%
t)-S^{x^{\prime }}\sin (\frac{\gamma B}{\hbar }t).  \label{gauge1}
\end{equation}%
Thus the Landau-Lifshitz equation (\ref{magnet}) in the rotating coordinate
system can be written as 
\begin{eqnarray}
\frac{\partial }{\partial t}S^{x} &=&S^{y}\frac{\partial ^{2}}{\partial z^{2}%
}S^{z}-S^{z}\frac{\partial ^{2}}{\partial z^{2}}S^{y}-16\rho ^{2}S^{y}S^{z},
\notag \\
\frac{\partial }{\partial t}S^{y} &=&S^{z}\frac{\partial ^{2}}{\partial z^{2}%
}S^{x}-S^{x}\frac{\partial ^{2}}{\partial z^{2}}S^{z}+16\rho ^{2}S^{z}S^{x},
\notag \\
\frac{\partial }{\partial t}S^{z} &=&S^{x}\frac{\partial ^{2}}{\partial z^{2}%
}S^{y}-S^{y}\frac{\partial ^{2}}{\partial z^{2}}S^{x},  \label{magnet2R}
\end{eqnarray}%
where $\rho ^{2}=\lambda ^{ld}/(4\lambda )$, and the superscript ( $^{\prime
}$\ ) is omitted for the sake of pithiness. The dimensionless time $t$ and
coordinate $z$ in Eq. (\ref{magnet2R}) are scaled in unit $2\lambda /\hbar $
and $a$ respectively, where $a$ denotes the lattice constant. Also, the
terms including the external magnetic field in Eq. (\ref{magnet}) have been
eliminated with the help of the transformation.

\section{One-soliton solution of spinor BEC in an optical lattice}

The equation (\ref{magnet2R}) has a form of the Landau-Lifshitz (LL) type
which is similar to the LL equation for a spin chain with an easy plane
anisotropy \cite{Note1}. By introducing a particular parameter Huang \cite%
{Huang} showed that the Jost solutions can be generated and the Lax
equations are satisfied, and moreover constructed Darboux transformation
matrices. An explicit expression of the one-soliton solution in terms of
elementary functions of $z$\ and $t$\ was reported. Here using the similar
method in Ref. \cite{Ablowitz,Huang} we obtain both the one-and two-soliton
solutions (for detail see the appendix) denoted by $\mathbf{S}(n)$\ with $%
n=1,2$ of Eq. (\ref{magnet2R})\textbf{\ }in the following form:

\begin{eqnarray}
S_{n}^{x} &=&1-\frac{1}{\Lambda _{n}}\left( \chi _{2,n}+2\chi _{3,n}\sin
^{2}\Phi _{n}\right) ,  \notag \\
S_{n}^{y} &=&\frac{-1}{\Lambda _{n}}\left( \chi _{1,n}\eta _{n}\cosh \Theta
_{n}\sin \Phi _{n}+\chi _{2,n}\allowbreak \sinh \Theta _{n}\cos \Phi
_{n}\right) ,  \notag \\
S_{n}^{z} &=&\frac{1}{\Lambda _{n}}\left( \chi _{1,n}\cosh \Theta _{n}\cos
\Phi _{n}+\chi _{2,n}\eta _{n}\sinh \Theta _{n}\sin \Phi _{n}\right) ,
\label{solitonR}
\end{eqnarray}%
where the parameters in the solution are defined by 
\begin{eqnarray}
\Lambda _{n} &=&\cosh ^{2}\Theta _{n}+\chi _{3,n}\sin ^{2}\Phi _{n},  \notag
\\
\Theta _{n} &=&2\kappa _{4,n}\left( z-V_{n}t-z_{n}\right) ,  \notag \\
\Phi _{n} &=&2\kappa _{3,n}z-\Omega _{n}t+\phi _{n},  \notag \\
V_{n} &=&2\left( \kappa _{1,n}+\frac{\kappa _{3,n}}{\kappa _{4,n}}\kappa
_{2,n}\right) ,  \notag \\
\Omega _{n} &=&4\left( \kappa _{1,n}\kappa _{3,n}-\kappa _{2,n}\kappa
_{4,n}\right) ,  \label{para1R}
\end{eqnarray}%
with $\kappa _{1,n}=\mu _{n}(1+\rho ^{2}/|\zeta _{n}|^{2})$, $\kappa
_{2,n}=\nu _{n}(1-\rho ^{2}/|\zeta _{n}|^{2})$, $\kappa _{3,n}=\mu
_{n}(1-\rho ^{2}/|\zeta _{n}|^{2})$, $\kappa _{4,n}=\nu _{n}(1+\rho
^{2}/|\zeta _{n}|^{2})$, $\eta _{n}=(|\zeta _{n}|^{2}+\rho ^{2})/(|\zeta
_{n}|^{2}-\rho ^{2})$, $\chi _{1,n}=\left( 2\mu _{n}\nu _{n}\right) /|\zeta
_{n}|^{2}$, $\chi _{2,n}=\left( 2\nu _{n}^{2}\right) /|\zeta _{n}|^{2}$, and 
$\chi _{3,n}=\left( 4\rho ^{2}\nu _{n}^{2}\right) /(|\zeta _{n}|^{2}-\rho
^{2})^{2}$. The one-soliton solution, namely $\mathbf{S}(1)$ , is simply%
\textbf{\ }that 
\begin{equation}
S^{x}(1)=S_{1}^{x};\quad S^{y}(1)=S_{1}^{y};\quad S^{z}(1)=S_{1}^{z}.
\label{onesoliton}
\end{equation}%
The parameter $V_{1}$\ denotes the velocity of envelope motion of the
magnetic soliton. The real constants $z_{1}$\ and $\phi _{1}$\ represent the
center position and the initial phase respectively. The parameter $\zeta
_{1}=\mu _{1}+i\nu _{1}$\ is eigenvalue with $\mu _{1}$, $\nu _{1}$\ being
the real and imaginary parts. The one-soliton solution (\ref{onesoliton})
describes a spin precession characterized by four real parameters: velocity $%
V_{1}$, phase $\Phi _{1}$, the center coordinate of the solitary wave $z_{1}$%
\ and initial phase $\phi _{1}$. From the one-soliton solution we obtain the
properties of the soliton: (i) both the amplitude and the size of the
soliton vary with time periodically, as shown in figure 1, in which we
demonstrate graphically the dynamics of soliton with\ the parameters chosen
as $\mu _{1}=0.45$\ , $\nu _{1}=0.7$, $\rho ^{2}=0.375$, $z_{1}=-14$\ and $%
\phi _{1}=0.5$. This property\textbf{\ }is resulted from the term $\rho $ in
Eq. (\ref{magnet2R}) which is determined by the light-induced dipole-dipole
interaction. This significant observation from the one-soliton solution
shows that the time-oscillation of the amplitude and the size of soliton can
be controlled in practical experiment by adjusting of the light-induced
dipole-dipole interaction. (ii) the magnetization defined by\textbf{\ }$%
M_{3}=\int_{-\infty }^{\infty }dz(1-S_{1}^{x})$ varies with time
periodically as shown in figure 2. These properties are similar to that of
the Heisenberg spin chain with an easy plane anisotropy where the dipolar
coupling is typically several orders of magnitude weaker than the exchange
coupling and thus would correspond to Curie temperatures much below the
observed values. Hence the contribution of the dipolar coupling to the spin
wave can be neglected in practice. However, for the spinor BEC in the
optical lattice the exchange interaction is absent and the individual spin
magnets are coupled by the magnetic and the light-induced dipole-dipole
interactions. Due to the large number of atoms $N$ at each lattice site,
these site to site interactions, despite the large distance between sites,
explain the natural existence of magnetic soliton which agrees with the
results in Refs. \cite{Pu,Zhang}.

To see closely the physical significance of one-soliton solution, it is
helpful to show the parameter-dependence of Euler angles of the
magnetization vector which in a spherical coordinate is 
\begin{equation}
S_{1}^{x}(z,t)=\cos \theta ,\text{ }S_{1}^{y}+iS_{1}^{z}=\sin \theta \exp
(i\varphi ).  \label{polar}
\end{equation}
From Eqs. (\ref{solitonR}) and (\ref{onesoliton}) we find 
\begin{eqnarray}
\cos \theta &=&1-\frac{\frac{2\nu _{1}^{2}}{\left\vert \zeta _{1}\right\vert
^{2}}+2\chi _{3,1}\sin ^{2}\Phi _{1}}{\cosh ^{2}\left[ \digamma
_{1}^{-1}\left( z-V_{1}t-z_{1}\right) \right] +\chi _{3,1}\sin ^{2}\Phi _{1}}%
,  \notag \\
\varphi &=&\frac{\pi }{2}+\arctan \left( \eta _{1}\tan \Phi _{1}\right)
+\arctan \left( \tanh \Theta _{1}\right) ,  \label{polar1R}
\end{eqnarray}
where $\digamma _{1}=1/(2\kappa _{4,1})$ and the maximal amplitude $%
A_{M}=2\left( \nu _{1}^{2}/\left\vert \zeta _{1}\right\vert ^{2}+\left\vert
\chi _{3,1}\right\vert \right) $. When $\left\vert \zeta _{1}\right\vert
^{2}>>\rho ^{2}$, the phase $\varphi $ can be rewritten as 
\begin{equation}
\varphi =\frac{\pi }{2}+\phi _{1}+k_{1}z-\Omega _{1}t+\arctan \left( \tanh
\Theta _{1}\right) ,
\end{equation}
where the wave number $k_{1}=2\kappa _{3,1}$ and the frequency of
magnetization precession $\Omega _{1}$ are related by the dispersion law 
\begin{equation}
\Omega _{1}=k_{1}\left( k_{1}+4\rho ^{2}\mu _{1}/\left\vert \zeta
_{1}\right\vert ^{2}\right) -4\kappa _{2,1}\kappa _{4,1}.  \label{polar4}
\end{equation}
We also see that the position of minimum of energy spectrum $\Omega _{1,\min
}=0$ is located at $k_{1\min }=(\sqrt{\left( 2\rho ^{2}\mu _{1}\right)
^{2}+4\kappa _{2,1}\kappa _{4,1}\left\vert \zeta _{1}\right\vert ^{4}}-2\rho
^{2}\mu _{1})/\left\vert \zeta _{1}\right\vert ^{2}$.

If amplitude $A_{M}$ approaches zero, namely $\nu _{1}\rightarrow 0$, the
parameter $\digamma _{1}$ diverges and Eq. (\ref{polar1R}) takes the
asymptotic form 
\begin{equation}
\cos \theta \rightarrow 1,\text{ }\varphi \rightarrow \frac{\pi }{2}+\phi
_{1}+k_{1}z-\Omega _{1}t,
\end{equation}
indicating a small linear solution of magnon. In this case the dispersion
law reduces to $\Omega _{1}=k_{1}\left( k_{1}+4\rho ^{2}\mu _{1}/\left\vert
\zeta _{1}\right\vert ^{2}\right) $.

\section{Elastic soliton collision for spinor BEC in an optical lattice}

The magnetic soliton collision in spinor BECs is an interesting phenomenon
in spin dynamics. Here in terms of a Darboux transformation we first of all
give the two-soliton solution (for detail see the appendix) of\textbf{\ }Eq.
(\ref{magnet2R}) 
\begin{eqnarray}
S^{x}\left( 2\right) &=&S_{1}^{x}S_{2}^{x}+R_{3}S_{1}^{y}+R_{5}S_{1}^{z}, 
\notag \\
S^{y}\left( 2\right) &=&S_{1}^{x}S_{2}^{y}+R_{4}S_{1}^{y}+R_{6}S_{1}^{z}, 
\notag \\
S^{z}\left( 2\right) &=&S_{1}^{x}S_{2}^{z}+R_{1}S_{1}^{y}+R_{2}S_{1}^{z},
\label{twosoliton1R}
\end{eqnarray}%
where $S_{n}^{x}$, $S_{n}^{y}$ and $S_{n}^{z}$ $\left( n=1,2\right) $ are
defined in Eq. (\ref{solitonR}) and $R_{j}$ $\left( j=1,2,...6\right) $ take
form as follows: 
\begin{eqnarray}
R_{1} &=&\frac{1}{\Lambda _{2}}\left( \chi _{1,2}\cosh \Theta _{2}\sinh
\Theta _{2}-\chi _{2,2}\eta _{2}\cos \Phi _{2}\sin \Phi _{2}\right) ,  \notag
\\
R_{2} &=&1-\frac{\chi _{2,2}}{\Lambda _{2}}\left( \cosh ^{2}\Theta _{2}-\sin
^{2}\Phi _{2}\right) ,  \notag \\
R_{3} &=&\frac{1}{\Lambda _{2}}\left( \chi _{1,2}\eta _{2}\cosh \Theta
_{2}\sin \Phi _{2}-\chi _{2,2}\sinh \Theta _{2}\cos \Phi _{2}\right) , 
\notag \\
R_{4} &=&1-\frac{1}{\Lambda _{2}}\left[ \chi _{2,2}\sinh ^{2}\Theta
_{2}+\left( \chi _{2,2}+2\chi _{3,2}\right) \sin ^{2}\Phi _{2}\right] , 
\notag \\
R_{5} &=&\frac{1}{\Lambda _{2}}(\chi _{2,2}\eta _{2}\sin \Phi _{2}\sinh
\Theta _{2}-\chi _{1,2}\cosh \Theta _{2}\cos \Phi _{2}),  \notag \\
R_{6} &=&\frac{-1}{\Lambda _{2}}\allowbreak (\chi _{2,2}\eta _{2}\sin \Phi
_{2}\cos \Phi _{2}+\chi _{1,2}\cosh \Theta _{2}\sinh \Theta _{2}),
\label{twosoliton2R}
\end{eqnarray}%
where $\Theta _{2}$, $\Phi _{2}$, $\Omega _{2}$, $\Lambda _{2}$, $\eta _{2}$
and $\chi _{m,2}$ $\left( m=1,2,3\right) $ are defined in Eq. (\ref{para1R}%
). The solution (\ref{twosoliton1R}) describes a general elastic scattering
process of two solitary waves with different center velocities $V_{1}$ and $%
V_{2}$, different phases $\Phi _{1}$ and $\Phi _{2}$. Before collision, they
move towards each other, one with velocity $V_{1}$ and shape variation
frequency $\Omega _{1}$ and the other with $V_{2}$ and $\Omega _{2}$. In
order to understand the nature of two-soliton interaction, we analyze the
asymptotic behavior of two-soliton solution (\ref{twosoliton1R}).
Asymptotically, the two-soliton waves (\ref{twosoliton1R}) can be written as
a combination of two one-soliton waves (\ref{onesoliton}) with different
amplitudes and phases. The asymptotic form of two-soliton solution in limits 
$t\rightarrow -\infty $ and $t\rightarrow \infty $ is similar to that of the
one-soliton solution (\ref{onesoliton}). In order to analyze the asymptotic
behavior of two-soliton solutions (\ref{twosoliton1R}) we\textbf{\ }show
first of all the asymptotic behavior of $S_{n}^{x}$, $S_{n}^{y}$, $%
S_{n}^{z}\left( n=1,2\right) $, and $R_{j}$($j=1,2,...6$) in the
corresponding limits $t\rightarrow \pm \infty $ from Eqs. (\ref{solitonR})
and (\ref{twosoliton2R}) 
\begin{eqnarray}
R_{1} &\rightarrow &\pm \chi _{1,2},\text{ \ }R_{2}\rightarrow 1-\chi _{2,2},%
\text{ \ }R_{3}\rightarrow 0,  \notag \\
R_{4} &\rightarrow &1-\chi _{2,2},\text{ \ }R_{5}\rightarrow 0,\text{ \ }%
R_{6}\rightarrow \mp \chi _{1,2},  \notag \\
S_{n}^{x} &\rightarrow &1,\text{ \ }S_{n}^{y}\rightarrow 0,\text{ \ }%
S_{n}^{z}\rightarrow 0,\text{as }t\rightarrow \pm \infty .
\end{eqnarray}%
Without loss of generality, we assume that $\kappa _{4,n}>0$ $\left(
n=1,2\right) $ and $V_{1}>V_{2}$ which corresponds to a head-on collision of
the solitons. For the above parametric choice, the variables $\Theta _{n}$($%
n=1,2$) for the two-soliton behave asymptotically as (i) $\Theta _{1}\sim 0$%
, $\Theta _{2}\sim \pm \infty $, as $t\rightarrow \pm \infty $; and (ii) $%
\Theta _{2}\sim 0$, $\Theta _{1}\sim \mp \infty $, as $t\rightarrow \pm
\infty $. This leads to the following asymptotic forms for the two-soliton
solution. (For the other choices of $\kappa _{4,n}$ and $V_{n}$, similar
analysis can be performed straightforwardly).

(i) Before collision, namely the case of limit $t\rightarrow -\infty $.

(a) Soliton 1 ($\Theta _{1} \sim 0$, $\Theta _{2} \rightarrow -\infty $). 
\begin{equation}
\left( 
\begin{array}{c}
S^{x}\left( 2\right) \\ 
S^{y}\left( 2\right) \\ 
S^{z}\left( 2\right)%
\end{array}
\right) \rightarrow \left( 
\begin{array}{c}
S_{1}^{x} \\ 
\sin \theta \cos \left( \varphi -\phi _{\Delta }\right) \\ 
\sin \theta \sin \left( \varphi -\phi _{\Delta }\right)%
\end{array}
\right) ,  \label{asym1a}
\end{equation}
where $\phi _{\Delta }=\arctan \left[ 2\mu _{2}\nu _{2}/\left( \mu
_{2}^{2}-\nu _{2}^{2}\right) \right]$ and the parameters $\theta $ and $%
\varphi $ are defined in Eq. (\ref{polar1R}).

(b) Soliton 2 ( $\Theta _{2} \sim 0$, $\Theta _{1}\rightarrow \infty $). 
\begin{equation}
\left( 
\begin{array}{c}
S^{x}\left( 2\right) \\ 
S^{y}\left( 2\right) \\ 
S^{z}\left( 2\right)%
\end{array}
\right) \rightarrow \left( 
\begin{array}{c}
S_{2}^{x} \\ 
S_{2}^{y} \\ 
S_{2}^{z}%
\end{array}
\right) ,  \label{asym2a}
\end{equation}

(ii) After collision, namely the case of limit $t\rightarrow \infty $.

(a) Soliton 1 ($\Theta _{1} \sim 0$, $\Theta _{2} \rightarrow \infty $). 
\begin{equation}
\left( 
\begin{array}{c}
S^{x}\left( 2\right) \\ 
S^{y}\left( 2\right) \\ 
S^{z}\left( 2\right)%
\end{array}
\right) \rightarrow \left( 
\begin{array}{c}
S_{1}^{x} \\ 
\sin \theta \cos \left( \varphi +\phi _{\Delta }\right) \\ 
\sin \theta \sin \left( \varphi +\phi _{\Delta }\right)%
\end{array}
\right) ,  \label{asym1b}
\end{equation}

(b) Soliton 2 ( $\Theta _2\sim 0$, $\Theta _1\rightarrow -\infty $). 
\begin{equation}
\left( 
\begin{array}{c}
S^x\left( 2\right) \\ 
S^y\left( 2\right) \\ 
S^z\left( 2\right)%
\end{array}
\right) \rightarrow \left( 
\begin{array}{c}
S_2^x \\ 
S_2^y \\ 
S_2^z%
\end{array}
\right) .  \label{asym2b}
\end{equation}
here for the expressions of solitons before and after collision, $S_n^x$, $%
S_n^y$ and $S_n^z$ $\left( n=1,2\right) $ are defined in Eq. (\ref{solitonR}%
). Analysis reveals that there is no amplitude exchange among three
components $S^x$, $S^y$ and $S^z$ for soliton 1 and soliton 2 during
collision. However, from Eqs. (\ref{asym1a}) and (\ref{asym1b}) one can see
that there is a phase exchange $2\phi _\Delta $ between two components $S^y$
and $S^z$ for soliton 1 during collision. This elastic collision between two
magnetic solitons in the optical lattice is different from that of coupled
nonlinear Schr\"{o}dinger equations \cite{Kanna}. It shows that the
information held in each soliton will almost not be disturbed by each other
in soliton propagation. These properties may have potential application in
future quantum communication. It should be noted that the inelastic
collision may appear if the influence of higher-order terms in Eq. (\ref%
{hamilton2}) is considered.

\section{Conclusion}

Magnetic soliton dynamics of spinor BECs in an optical lattice is studied in
terms of a modified Landau-Lifshitz equation which is derived from the
effective Hamiltonian of a pseudospin chain. The soliton solutions are
obtained analytically and the elastic collision of two solitons is
demonstrated. The significant observation is that time-oscillation of the
soliton amplitude and size can be controlled by adjusting of the
light-induced dipole-dipole interactions.

It should be interesting to discuss how to create the magnetic soliton and
how to detect such magnetic soliton in experiment. In the previous work \cite%
{Miesner} using Landau-Zener rf-sweeps at high fields (30 G) a condensate
was prepared in the hyperfine state $\left\vert f=1,m_{f}=0\right\rangle $,
i.e. the the ground state of the spinor BECs. Then the atoms of the ground
state can be excited to the hyperfine state $\left\vert f=1,m_{f}=\pm
1\right\rangle $ by laser light experimentally. Therefore the excited state
of the spinor BECs, i.e. the magnetic soliton can be created. As can be seen
from Fig. 1, the spatial-temporal spin variations in the soliton state are
significant. This makes it possible to take a direct detection of the
magnetic soliton of spinor BECs. By counting the difference numbers of the
population between the spin $+1$\ and $-1$\ Zeeman sublevel, the average of
spin component $<S^{z}>$\ is measured directly. While transverse components
can be measured by use of a short magnetic pulse to rotate the transverse
spin component to the longitudinal direction. Any optical or magnetic method
which can excite the internal transitions between the atomic Zeeman
sublevels can be used for this purpose. In current experiments in optical
lattices, the lattice number is in the range of $10$-$100$, and each lattice
site can accommodate a few thousand atoms. This leads to a requirement for
the frequency measurement precision of about 10-100 kHz. This is achievable
with current techniques. We can also see that the detection of the magnetic
soliton of the spinor BECs is different from that of the Heisenberg spin
chain.

The magnetic soliton of spinor BECs in an optical lattice is mainly caused
by the magnetic and the light-induced dipole-dipole interactions between
different lattice sites. Since these long-range interactions are highly
controllable the spinor BECs in optical lattice which is an exceedingly
clean system can serve as a test ground to study the static and dynamic
aspects of soliton excitations.

\section{Acknowledgement}

This work was supported by the NSF of China under Grant Nos. 10475053,
60490280, 90406017 and provincial overseas scholar foundation of Shanxi.

\section{Appendix}

The corresponding Lax equations for the Eq. (\ref{magnet2R}) are written as

\begin{equation}
\partial _{z}G\left( z,t\right) =LG\left( z,t\right) ,\partial _{t}G\left(
z,t\right) =MG\left( z,t\right) ,  \label{lax1}
\end{equation}%
where 
\begin{eqnarray}
L &=&-i\epsilon S^{z}\sigma _{3}-i\varsigma \left( S^{x}\sigma
_{1}+S^{y}\sigma _{2}\right) ,  \notag \\
M &=&i2\varsigma ^{2}S^{z}\sigma _{3}+i2\varsigma \epsilon \left(
S^{x}\sigma _{1}+S^{y}\sigma _{2}\right) -i\varsigma \left( S^{y}\partial
_{z}S^{z}-S^{z}\partial _{z}S^{y}\right) \sigma _{1}  \notag \\
&&-i\varsigma \left( S^{z}\partial _{z}S^{x}-S^{x}\partial _{z}S^{z}\right)
\sigma _{2}-i\epsilon \left( S_{1}\partial _{z}S^{y}-S^{y}\partial
_{z}S^{x}\right) \sigma _{3},  \label{lax2}
\end{eqnarray}%
here $\sigma _{j}$($j=1,2,3$) is Pauli matrix, the parameters $\epsilon $
and $\varsigma $ satisfy the relation $\epsilon ^{2}=\varsigma ^{2}+4\rho
^{2}$. Thus Eq. (\ref{magnet2R}) can be recovered from the compatibility
condition $\partial _{t}L-\partial _{x}M+[L,M]=0$. We introduce an auxiliary
parameter $q$ such that 
\begin{equation}
\epsilon =2\rho \frac{q+q^{-1}}{q-q^{-1}},\varsigma =2\rho \frac{2}{q-q^{-1}}%
,  \label{lax2a}
\end{equation}%
\textbf{\ }and the complex parameter is defined by\textbf{\ }$q=\left( \zeta
+\rho \right) /\left( \zeta -\rho \right) $.

It is easily to see that $S_{0}=\left( 1,0,0\right) $ is a simplest solution
of Eq. (\ref{magnet2R}). Under this condition the corresponding Jost
solution of Eq. (\ref{lax1}) can be written as

\begin{equation}
G_{0}=U\exp \left\{ -i\varsigma \left( z-2\epsilon t\right) \sigma
_{3}\right\} ,  \label{lax2c}
\end{equation}%
where $U=\frac{1}{2}\left\{ I-i\left( \sigma _{1}+\sigma _{2}+\sigma
_{3}\right) \right\} $ with $I$ denoting unit matrix. In the following we
construct the Darboux matrix $D_{n}\left( q\right) $ by using the following
recursion relation 
\begin{equation}
G_{n}\left( q\right) =D_{n}\left( q\right) G_{n-1}\left( q\right) ,\text{ }%
n=1,2,3,...,  \label{lax3}
\end{equation}%
where $D_{n}\left( q\right) $ has poles. Since

\begin{equation}
\epsilon \left( -\overline{q}\right) =\overline{\epsilon \left( q\right) }%
,\varsigma \left( -\overline{q}\right) =-\overline{\varsigma \left( q\right) 
},L\left( -\overline{q}\right) =\sigma _{1}\overline{L\left( q\right) }%
\sigma _{1},M\left( -\overline{q}\right) =\sigma _{1}\overline{M\left(
q\right) }\sigma _{1},  \label{lax4}
\end{equation}%
we then have 
\begin{equation}
G_{0}\left( -\overline{q}\right) =-i\sigma _{1}\overline{G_{0}\left(
q\right) },G_{n}\left( -\overline{q}\right) =-i\sigma _{1}\overline{%
G_{n}\left( q\right) },D_{n}\left( -\overline{q}\right) =\sigma _{1}%
\overline{D_{n}\left( q\right) }\sigma _{1},  \label{lax5}
\end{equation}%
where the overbar denotes complex conjugate. Suppose\textbf{\ }that $q_{n}$
is a simple pole of $D_{n}\left( q\right) ,$ then $-\overline{q}_{n}$ is
also a pole of $D_{n}\left( q\right) $. If $D_{n}\left( q\right) $ has only
these two simple poles we have 
\begin{eqnarray}
D_{n}\left( q\right) &=&C_{n}P_{n}\left( q\right) ,  \label{darboux1a} \\
P_{n}\left( q\right) &=&I+\frac{q_{n}-\overline{q}_{n}}{q-q_{n}}P_{n}+\frac{%
q_{n}-\overline{q}_{n}}{q+\overline{q}_{n}}\widetilde{P}_{n},
\label{darboux1b}
\end{eqnarray}%
where $C_{n}$, $P_{n}$, and $\widetilde{P}_{n}$\ are $2\times 2$\ matrix
independent of\textbf{\ }$q$, the terms $\left( q_{n}-\overline{q}%
_{n}\right) C_{n}P_{n}$\textbf{\ }and\textbf{\ }$\left( q_{n}-\overline{q}%
_{n}\right) C_{n}\widetilde{P}_{n}$\textbf{\ }are\textbf{\ }residues at%
\textbf{\ }$q_{n}$\textbf{\ }and\textbf{\ }$\overline{q}_{n}$, respectively.
From Eq (\ref{lax5}),\textbf{\ }we have 
\begin{equation}
C_{n}=\sigma _{1}\overline{C}_{n}\sigma _{1},\widetilde{P}_{n}=\sigma _{1}%
\overline{P_{n}}\sigma _{1}.  \label{darboux2}
\end{equation}%
From Eqs. (\ref{lax2}) and (\ref{lax2c}) we see that 
\begin{equation}
L(q)=-L^{\dag }\left( \overline{q}\right) ,M(q)=-M^{\dag }\left( \overline{q}%
\right) ,G_{0}^{-1}(q)=G_{0}^{\dag }\left( \overline{q}\right) ,
\label{darboux3}
\end{equation}%
and hence we have 
\begin{equation}
G_{n}^{-1}(q)=G_{n}^{\dag }\left( \overline{q}\right)
,D_{n}^{-1}(q)=D_{n}^{\dag }\left( \overline{q}\right) =P_{n}^{\dag }\left( 
\overline{q}\right) C_{n}^{\dag }.  \label{darboux4}
\end{equation}%
Since $D_{n}\left( q\right) D_{n}^{-1}\left( q\right) =D_{n}\left( q\right)
D_{n}^{-1}\left( \overline{q}\right) =I$, it has no poles. Then we obtain 
\begin{equation}
P_{n}P_{n}^{\dag }\left( \overline{q}_{n}\right) =0,\text{ }P_{n}\left(
I-P_{n}^{\dag }+\frac{\overline{q}_{n}-q_{n}}{2q_{n}}\widetilde{P}_{n}^{\dag
}\right) =0,  \label{darboux5}
\end{equation}%
which shows the degeneracy of $P_{n}$. One can write $P_{n}=(%
\begin{array}{ll}
g_{n} & w_{n}%
\end{array}%
)^{T}(%
\begin{array}{ll}
\Upsilon _{n} & \xi _{n}%
\end{array}%
)$ where the superscript $T$ means transpose. Substituting this expression
into (\ref{darboux5}) we obtain 
\begin{eqnarray}
P_{n}\left( q\right) &=&\frac{1}{\Delta _{n}\left( q-q_{n}\right) \left( q+%
\overline{q}_{n}\right) }\left( 
\begin{array}{cc}
\overline{q}_{n}\left\vert \Upsilon _{n}\right\vert ^{2}+q_{n}\left\vert \xi
_{n}\right\vert ^{2} & 0 \\ 
0 & q_{n}\left\vert \Upsilon _{n}\right\vert ^{2}+\overline{q}_{n}\left\vert
\xi _{n}\right\vert ^{2}%
\end{array}%
\right)  \notag \\
&&\times \left\{ q^{2}\left( 
\begin{array}{cc}
q_{n}\left\vert \Upsilon _{n}\right\vert ^{2}+\overline{q}_{n}\left\vert \xi
_{n}\right\vert ^{2} & 0 \\ 
0 & \overline{q}_{n}\left\vert \Upsilon _{n}\right\vert ^{2}+q_{n}\left\vert
\xi _{n}\right\vert ^{2}%
\end{array}%
\right) \right. +q\left( q_{n}^{2}-\overline{q}_{n}^{2}\right) \left( 
\begin{array}{cc}
0 & \overline{\Upsilon }_{n}\xi _{n} \\ 
\overline{\xi }_{n}\Upsilon _{n} & 0%
\end{array}%
\right)  \notag \\
&&-\left. \left\vert q_{n}\right\vert ^{2}\left( 
\begin{array}{cc}
q_{n}\left\vert \xi _{n}\right\vert ^{2}+\overline{q}_{n}\left\vert \Upsilon
_{n}\right\vert ^{2} & 0 \\ 
0 & q_{n}\left\vert \Upsilon _{n}\right\vert ^{2}+\overline{q}_{n}\left\vert
\xi _{n}\right\vert ^{2}%
\end{array}%
\right) \right\} ,  \label{darboux6}
\end{eqnarray}%
where 
\begin{equation}
\Delta _{n}=\left\vert q_{n}\right\vert ^{2}\left( \left\vert \Upsilon
_{n}\right\vert ^{2}+\left\vert \xi _{n}\right\vert ^{2}\right)
^{2}+\left\vert \overline{q}_{n}-q_{n}\right\vert ^{2}\left\vert \Upsilon
_{n}\right\vert ^{2}\left\vert \xi _{n}\right\vert ^{2}.  \label{darboux7}
\end{equation}%
To determine $\xi _{n}$ and $\Upsilon _{n},$ we substitute (\ref{lax3}) into
(\ref{lax1}) and take the limit $q\rightarrow q_{n}$ and then obtain 
\begin{eqnarray}
\partial _{z}D_{n}\left( q\right) &=&L_{n}\left( q\right) D_{n}\left(
q\right) -D_{n}\left( q\right) L_{n-1}\left( q\right) ,  \notag \\
\partial _{t}D_{n}\left( q\right) &=&M_{n}\left( q\right) D_{n}\left(
q\right) -D_{n}\left( q\right) M_{n-1}\left( q\right) .  \label{darboux8}
\end{eqnarray}%
Because of the degeneracy of $P_{n},$ the second factor of the right-hand
sides of Eq. (\ref{darboux8}), namely, ($%
\begin{array}{ll}
\Upsilon _{n} & \xi _{n}%
\end{array}%
)G_{n-1}\left( q_{n}\right) $ must appear in the left-hand side in its
original form and, hence, it is independent of $z$ and $t$. We simply let 
\begin{equation}
\left( 
\begin{array}{ll}
\Upsilon _{n} & \xi _{n}%
\end{array}%
\right) =\left( 
\begin{array}{ll}
b_{n} & 1%
\end{array}%
\right) G_{n-1}^{-1}\left( q_{n}\right) ,  \label{darboux9}
\end{equation}%
here $b_{n}$ is a constant. Hence, the Darboux matrices $D_{n}\left(
q\right) $ have been determined recursively, except for $C_{n}.$ By a simple
algebraic procedure,\textbf{\ }it is seen that $\Delta _{n}$ is always
non-vanishing regardless of the values $z$ and $t.$ This shows the
regularity of $P_{n}$ and then $P_{n}\left( q\right) $. In the limit as $%
q\rightarrow 1$, from Eq. (\ref{lax2a}) we have 
\begin{equation*}
\epsilon \left( q\right) ,\varsigma \left( q\right) \rightarrow 2\rho \frac{1%
}{q-1}+O\left( 1\right) ,
\end{equation*}%
and then from Eq. (\ref{darboux8}) we obtain 
\begin{equation}
S\left( n\right) \cdot \sigma =D_{n}\left( 1\right) \left[ S\left(
n-1\right) \cdot \sigma \right] D_{n}^{\dag }\left( 1\right) ,n=1,2,3,....
\label{solution1}
\end{equation}%
Considering the Eq. (\ref{darboux3}) and Eq. (\ref{darboux4}) we get 
\begin{equation}
C_{n}C_{n}^{\dag }=I,
\end{equation}%
which shows that the matrix $C_{n}$ is diagonal with the help of the Eq. (%
\ref{darboux2}) and 
\begin{equation}
\left( C_{n}\right) _{11}=\left( \overline{C_{n}}\right) _{22},\left\vert
\left( C_{n}\right) _{11}\right\vert =1,
\end{equation}%
then we can write $C_{n}=\exp \left( i\omega _{n}\sigma _{3}/2\right) $
which is real and characterizes the rotation-angle of spin in the $xy$%
-plane. It is necessary to mention that $\omega _{n}$ may he dependent on $z$
and $t$. To determine $\omega _{n}$ one must examine the Lax equations
carefully. Since $\exp \left( i\omega _{n}\sigma _{3}/2\right) $ denotes a
rotation around the $z$-axis, it does not affect the value of $S^{z}$.
Substituting (\ref{darboux1a}) into (\ref{darboux8}) and taking the limits
as $q\rightarrow \infty $ and $q\rightarrow 0$ respectively, we obtain 
\begin{eqnarray*}
\partial _{z}\left\{ C_{n}P_{n}\left( 0\right) \right\} &=&L_{n}\left(
q\right) \left\{ C_{n}P_{n}\left( 0\right) \right\} -\left\{
C_{n}P_{n}\left( 0\right) \right\} L_{n-1}\left( q\right) , \\
\partial _{z}\left\{ C_{n}\right\} &=&-i2\rho S^{z}\left( n\right) \sigma
_{3}\left\{ C_{n}\right\} +\left\{ C_{n}\right\} i2\rho S^{z}\left(
n-1\right) \sigma _{3}.
\end{eqnarray*}%
Comparing these two equations, we derive 
\begin{equation}
C_{n}=\left( \Delta _{n}\right) ^{-\frac{1}{2}}\left( 
\begin{array}{cc}
q_{n}\left\vert \Upsilon _{n}\right\vert ^{2}+\overline{q}_{n}\left\vert \xi
_{n}\right\vert ^{2} & 0 \\ 
0 & \overline{q}_{n}\left\vert \Upsilon _{n}\right\vert ^{2}+q_{n}\left\vert
\xi _{n}\right\vert ^{2}%
\end{array}%
\right) .  \label{darboux10}
\end{equation}%
The Eq. (\ref{darboux9}) gives 
\begin{equation}
\Upsilon _{n}=f_{n}+if_{n}^{-1},\xi _{n}=f_{n}-if_{n}^{-1},
\label{darboux11}
\end{equation}%
where 
\begin{equation*}
f_{n}^{2}=\exp \left( -\Theta _{n}+i\Phi _{n}\right) ,
\end{equation*}%
here the parameters $\Theta _{n}$ and $\Phi _{n}$ are defined in Eq. (\ref%
{para1R}). Setting $n=1$ and substituting the Eqs. (\ref{darboux1a}), (\ref%
{darboux6}), (\ref{darboux10}), and (\ref{darboux11}) into (\ref{solution1})
we can obtain the one-soliton solution (\ref{onesoliton}). Setting $n=2$ and
with the similar procedure the expression of the two-soliton solution (\ref%
{twosoliton1R}) is derived.

Figure Captions\newline

Figure 1 The amplitude and size of soliton in Eq. (\ref{onesoliton}) vary
periodically with time, where $\mu _{1}=0.45$, $\nu _{1}=0.7$, $z_{1}=-14$, $%
\phi _{1}=0.5$, $\rho ^{2}=0.375$. The unit for time $t$\ is $2\lambda
/\hbar $\ \ and $a$\ for space $z$.

Figure 2 The magnetization $M_{3}$ (the integral $\int_{-\infty }^{\infty
}dz(1-S_{1}^{x})$) vary with time periodically, where $\mu _{1}=0.45$, $\nu
_{1}=0.7$, $z_{1}=-14$, $\phi _{1}=0.5$, $\rho ^{2}=0.375$. The unit for
time $t$\ is $2\lambda /\hbar $\ \ and $a$\ for space $z$\textbf{.}

\end{document}